\documentstyle[aps,pre,epsfig]{revtex}
\begin{document}
\draft 
\title{First passage time exponent for higher-order random walks: Using
Levy flights}
\date{\today}
\author{J. M. Schwarz$^1$ and  Ron Maimon$^2$}
\address{$^1$Lyman Laboratory of Physics, Harvard University, Cambridge, 
Massachusetts 02138 and $^2$Neuman Laboratory, Cornell University, Ithaca,
New York 14850}
\maketitle

\begin{abstract}
We present a heuristic derivation 
of the first passage time exponent 
for the 
integral of a random walk [Y. G. Sinai, Theor. Math. Phys. {\bf 90}, 219
(1992)].  Building on this derivation, we 
construct an estimation scheme to 
understand the first passage time exponent 
for the integral of the integral
of a random walk, which is numerically observed to be 
$0.220\pm0.001$.  We discuss the
implications of this estimation scheme for the $n{\rm th}$ integral of a 
random walk.  For completeness, we also address the $n=\infty$ case.       
Finally, we explore an application of these processes to an 
extended, elastic object being pulled through a random potential by
a uniform applied force. In so doing, we
demonstrate a time reparameterization freedom in the Langevin
equation that maps nonlinear stochastic processes into linear ones.
\end{abstract} 


\def\half{{1\over 2}}
\def\third{{1\over 3}}
\def\fourth{{1\over 4}}
\def\fifth{{1\over 5}}
\def\sixth{{1\over 6}}
\def\eighth{{1\over 8}}
\def\twothirds{{2\over 3}}
\def\threefourths{{3\over 4}}
\def\twofifths{{2\over 5}}
\def\threefifths{{3\over 5}}
\def\fourfifths{{4 \over 5}}
\def\fivesixths{{5\over 6}}
\def\threehalves{{3\over 2}}
\def\fivefourths{{5\over 4}}

\def\ddt{{d\over dt}}
\def\ddx{{d\over dx}}
\def\ddxx{{d^2\over dx^2}}
\def\inverse#1{{1\over #1 }}
\def\invsqrt#1{{1\over \sqrt{#1}}}
\def\gaussian#1#2{ e^{- { {#1}^2 \over #2 } }}

\section{Introduction}

We investigate the general random walk $x(t)$ obeying the
equation of motion
\begin{equation}
{d^n x(t) \over dt^n}=\eta(t),
\end{equation}  
where $\eta$ is white noise with zero mean and 
unit variance and $x(0)=x_{o}$. 
   
Let us begin with $n=1$.  
The \emph{first passage time} is the time it takes for the walk to
reach zero. 
When the walk has no bias, as above, there is no
definite time to expect such an event 
and the distribution is
a power law.  To find the first passage time 
distribution $\rho(t)dt$, start an 
ensemble of random walkers
at $x_{o}>0$ and at time $t=0$. Whenever a random 
walk reaches zero, it is removed from the
ensemble. Let $P_{t}(x)$ be the number density of walks at time $t$
and positive $x$. $P_{t}(x)$ is a 
solution of the diffusion equation 
with absorbing boundary
condition, $P_{t}(0)=0$.  More precisely, 
\begin{eqnarray}
P_{t}(x)= \invsqrt{2\pi t} (\gaussian{(x-x_{o})}{2t} - 
\gaussian{(x+x_{o})}{2t}) \nonumber \\
\approx  2x_{o}\ddx( \invsqrt{2\pi t} 
\gaussian{x}{2t}) 
\end{eqnarray}
at long times.

Let $g(t)$ be the integral of this probability distribution over positive values of
$x$.  This is
the probability that a random walk at time $t$ has not crossed zero.
The first passage time distribution is then given by
\begin{equation}
\rho(t)dt = -\frac{dg}{dt}dt = {x_{o} \over \sqrt{2\pi t}} {dt \over t}\ . 
\end{equation}
Although we will not show it, this result is universal for all 
symmetric walks.  Any random walk which 
is equally likely to move forward as backward by a given amount has a first
passage time distribution with the same asymptotic form. This is
the Sparre-Anderson theorem [1]. Note that $x_{o}^2$ sets the time scale. 

The previous discussion is one of the few first passage time 
problems where an exact solution may be easily found
by solving a Fokker-Planck equation with absorbing boundaries.
Extensions of this method to more complicated random
walks, such as the second-order random walk described 
by
\begin{equation}
\frac{d^{2}x}{dt^2}=\eta(t), 
\end{equation}
exist [2].  However, we investigate the first passage time distribution 
for this walk, 
and for any $n$, in a different way.

At long times, the first passage time distribution for these processes 
is a power law,
\begin{equation}
\rho_{n}(t)dt\sim\frac{1}{t^{\beta_{n}}}\frac{dt}{t}\,.
\end{equation}
As computed above, $\beta_{1} = \half$. 
$\beta_{2}$ is known to be $\fourth$ [3].  
The others are most likely not exact fractions and are nontrivial critical
exponents of certain statistical models.  We first present numerical results for
$n=2,3,4,5$.  Then, after presenting a 
heuristic derivation of $\beta_2$, we make a quantitative estimate for 
the shift in first passage exponent
for $n\ge 3$.  Our analytical results will 
draw heavily from the theory of Levy flights [4].  We also 
address separately the $n=\infty$ limit [5].  

Finally, within zero temperature mean field theory, we 
demonstrate that the first passage time exponent in the $n=2$ case
is the avalanche size exponent for a dynamically elastic extended object, 
like a crack front or interface, being pulled through a random
medium by a uniform applied force at a special point in the parameter 
space [6].  Generically, for these
nonequilibrium systems 
there is a transition from an overall stationary phase to 
an overall moving phase as the applied force is increased toward
a critical value.  Increasing 
the applied force from the static side triggers
\emph{local} motion of the interface for some finite amount of time as long 
as the applied force is held fixed from the
time the toppling starts until the time the toppling stops.  The 
distribution of amount the interface has moved during
an ``avalanche'' event, i.e. the avalanche size, gives us 
information about the continuity (or discontinuity) of the depinning
transition and, therefore, has been a focus of study over the years [7].

\section{Levy Flights}

A \emph{Levy variable} $q_i$ is a random variable with a power law 
distribution for large $q$, 
\begin{equation}
P(q)dq\propto \frac{1}{q^{\beta}}
\frac{dq}{q} ,
\end{equation}
with $0<\beta<2$. The variance $<q^2>$ is infinite
for these distributions and $\beta$ is called the {\em Levy exponent} of $q$.

A \emph{Levy flight} is a random walk with each step a Levy variable. It is
the sum of many independently drawn Levy variables. Because of the infinite
variance, a Levy flight is an irregular walk dominated by the few largest 
jumps [4].

Let $L_{N}(Q)$ be the distribution of $Q=\sum_{i=1}^{N}q_i$.  
The Fourier transform $\tilde{L}_{N}(k)$ 
is the $N$th power of the Fourier
transform of $P(q)$. If $\tilde{P}(k)$ were twice differentiable at zero, then
$\tilde{L}_{N}(k)\sim (1-bk^2+ {\cal O}(k^3))^N\approx 
e^{-b N k^2}$ for $b$ real
and positive. This is the central limit theorem, and it applies when the
second moment of a distribution is finite. For a Levy variable, the
second moment of the distribution is infinite, and $\tilde{P}(k)$ has a cusp
at zero for $0<\beta<2$. The form of the cusp may be determined as follows:
for small $k$, the Fourier
transform is approximately the integral of the distribution over the first
wavelength $\lambda=\frac{2\pi}{k}$, or
\begin{equation}
\int_1^\lambda {C \over q^\beta} {dq \over q} =
1 - C'\frac{1}{\lambda^\beta} = 1 - C'' k^\beta.
\end{equation}
When there is a cusp, the limiting distribution is not a Gaussian, but has
the following form [4]
\begin{equation}
\tilde{L}_N(k)=e^{-bN|k|^\beta} \ \ \ k>0
\end{equation}
\[
\tilde{L}_N(k)=e^{-b^*N|k|^\beta} \ \ \ k<0 .
\]
By rescaling $k$, $b$ can be made into a pure phase.
For the case where $P(q)$ is symmetric about zero, the Fourier transform is
real and $b=1$.
It is clear that this distribution has the correct form near $k=0$, and
increasing $N$ is equivalent to a rescaling of $k$. The limiting 
distribution gets
wider without changing shape --- so it is a fixed point of convolution.

We will 
need one special nonsymmetric distribution, the distribution of a flight
composed of only positive steps. In this case, $L_N(Q)$ is zero for
$Q$ negative, therefore $\tilde{L}_{N}(k)$ must be analytic
in the lower half-plane. In other words, $\tilde{L}(k)$ for $k<0$ is
the analytic continuation of $\tilde{L}(k)$ for $k>0$.
As $k$ is rotated to $-k$ through negative imaginary values,
$k^\beta$  acquires a phase $e^{-\pi i \beta}$. We conclude that
\begin{equation}
b^* = b  e^{-i\pi \beta}
\end{equation}
\[
b=e^{i\pi  \beta /2} .
\] 
For the special case $\beta = \half$, $b=\frac{1+i}{\sqrt{2}}$.

And finally, if we were to compute the 
first passage time distribution for the process,
\begin{equation}
\frac{dx}{dt}=\xi(t),
\end{equation}
where $\xi(t)$ is Levy noise (a Levy variable symmetrically
distributed about zero), then we would find a first
passage time exponent of $\frac{1}{2}$.
Even though the Levy flight is irregular, the position does not cross 
zero any faster than it does in the nearest-neighbor random walk.
This is the Sparre-Anderson theorem once more, which we refer the reader
to Ref. [1] for the details.

\section{Numerical Results}

To obtain the first passage time exponent $\beta_n$ numerically
is not as easy as it might appear. Direct numerical integration of the
equation of motion becomes more cumbersome with increasing $n$.  
To efficiently simulate the equation, we have calculated the free-space 
propagation
kernels for $n=2,3,4,5$ (Appendix A).  In this context, the propagation
kernel is the probability distribution of $x(t_o + t_s)$, given
initial values $x(t_o)$, $x'(t_o)$, ... , $x^{(n-1)}(t_o)$
and final values $x'(t_o+t_s)$, ... $x^{(n-1)}(t_o+t_s)$.
We first generate the highest-order time derivative, $x^{(n-1)}(t_o+t_s)$, 
and then new values for each lower-order time derivative 
until $x(t_o+t_s)$ is updated.

The next time step $t_s$ is chosen so that the variance for the
next value of $x$ will be
smaller. The ratio of the new variance to the old is $\chi$.  $\chi$ 
should be small so that the walk does not become negative
then positive within one time step.  The probability of this occurring is 
exponentially small in the inverse of $\chi$.  
With this algorithm, if $x(t_o)$ is large, the time 
steps are large as well. The number
of time steps required for the simulation only grows logarithmically 
as a function of the first passage time.  
Near the end of the simulation, when $x$ is small, the time steps
become small as well, and accuracy is not sacrificed. 

The first passage time distributions for $n=2,3,4,5$ are shown in Fig. 1 
on a log-log scale.  Each plot contains 
approximately $10^7$ 
samples in bins of doubling size.  Table I contains 
the linear regression values of the exponents.  
The results are independent of $\chi$ over the range $[0.025,0.005]$, 
indicating that $\chi$ is small enough. 

\section{The First Passage Time Exponent for $\lowercase{n}=2$}

For the $n=2$ case, $x(t)$ is the integral of a random walk. 
In other words, the variable that executes the random walk is $x'(t)$,
the velocity.  To exploit this fact, we introduce a new time counter $i$
and divide the time axis into intervals $\Delta t_i$ between the points
where the velocity crosses zero.
Then the interval sizes $\Delta t_i$ 
are first passage times for
an ordinary random walk.

There is one complication.  Referring 
back to Eq. (3), the time scale until a zero-crossing is the initial value 
squared. So right after a zero-crossing, the initial value is zero. Since the
probability distribution is singular at zero, the random walk reaches zero
again instantly, and then
again, infinitely many times. This is a well known property of random
walks --- they jiggle about every value before moving on. Since this
affects only the distribution of the smallest times, 
we cut off the distribution
of times near zero by imagining the system on a lattice.
Now we have a finite and discrete set of time intervals $\Delta t_i$ with each 
interval distributed with
the Levy exponent $\beta_1=\half$.  These time intervals are independently
distributed because the velocity undergoes a simple random walk.

To compute the first passage time for the position, 
we must compute the total time $t=\sum_{i=1}^{N} \Delta t_i$ until 
the position becomes negative.  $N$ is the first zero-crossing of the
velocity which happens at a negative value of the position so we are actually 
slightly overestimating
$t$ by summing $N$ intervals.  We can control for this 
by summing $N-1$ intervals
instead, which would be a slight underestimate.  This estimate of $t$, be it 
over or under, should not affect the first passage time exponent since we
do not expect the last position step (corresponding to the 
partial time step) to be arbitrarily large.

To continue we need to know two things ---
1) the distribution of $N$, $f(N)$, and 
2) given $N$, the probability distribution of $t$.

The first problem is actually no problem at all.  Since the 
velocity is just as likely to go
up at the start of a time interval as it is to go down, the position 
is as likely to be positive as negative.  Therefore, the position is 
a symmetric walk on the zero-crossings (actually, a symmetric Levy flight with
independent steps distributed with Levy
exponent 1/3 using scaling arguments).
By the Sparre-Anderson theorem [1],
the distribution of the first-passage-$N$ is asymptotically the same
as for an unbiased, simple random walk, and
\begin{equation}
f(N)\sim {1 \over N^\threehalves}
\end{equation}
for large $N$.

And now for the second problem.  The distribution
of the sum of $N$ {\em independent} $\Delta t$'s, $L_N(t)dt$, is a Levy distribution.  The fact that the steps are distributed with Levy exponent $\beta_1=\half$ 
and that 
all of the $\Delta t$'s are positive fixes the Fourier transform of 
the Levy distribution.
The inverse Fourier transform can be computed
exactly in this case. For large t,
\begin{equation}
L_N(t)dt \sim \frac{N}{t^{1/2}}\gaussian{N}{2t}\frac{dt}{t}.
\end{equation}
This distribution is small
for $N^2>t$, and it may be approximated by ${N \over t^\half}{dt \over t}$ 
for $N^2<t$.  Using scaling, analogous results may be derived for
$\beta \ne \half$, where a closed form does not exist.

Now, the time intervals we are adding  
are not really independent.  
This is because we are restricting attention to Levy flights that end on 
the $N$th step and this an atypical sample of all Levy flights with 
$N$ steps.  However, the properties we actually 
use from the previous distribution 
 --- namely that it   
is small for $N^2>t$ and 
the appropriate power law for large $t$ --- are the scaling
laws for the sum of almost any collection of $N$ Levy variables, where $N$ is large.  So, 
although we do not prove it, we assume that these properties hold for the 
correct $L_N(t)dt$, or the distribution of the time elapsed for those Levy flights that end on the $N$th step.  This is why we refer to our derivation as heuristic.
 
All the ingredients are now in place for computing $\beta_2$. There is
a probability $f(N)$ for any value of $N$, and given $N$, we know the
probability distribution of $t$.  So to find the total distribution of
$t$, we sum up the conditional distributions $L_N(t)$ weighted by the
probability of $N$, or
\begin{equation}
\rho_2(t)dt= \sum_{N=0}^\infty f(N) L_N(t)dt\,.
\end{equation}
Approximating the sum with an integral and using the approximate 
scaling form for $L_N(t)$, 
\begin{equation}
\rho_2(t)dt\sim\int_{0}^{\infty}dN\,f(N)L_N(t)dt
\sim \int_{0}^{\sqrt{t}} {1\over N^\threehalves} {N\over t^\threehalves} dN . 
\end{equation}
We obtain the following asymptotic form,
\begin{equation}
\rho_2(t)dt= \inverse{t^{\fourth}}\frac{dt}{t}\,,
\end{equation}
so that $\beta_2=\fourth$. This result is unchanged when $N$ is shifted by one unit,
so that the last step is of no consequence for the exponent 
as there are typically many zero-crossings, making our derivation
self-consistent.  

Y. G. Sinai has rigorously computed $\beta_2$ [3]; we have used 
some similar methods.

\section{The first passage time exponent for $\lowercase{n}\ge 3$}

The next random walk we consider is the random surge process,
governed by
\begin{equation}
{d^3x \over dt^3}={da\over dt}=\eta(t).      
\end{equation}
The acceleration is now an ordinary random walk. After a time t, the 
acceleration, velocity and position scale as
$a_{\rm typ}\sim t^{\half}$, $v_{\rm typ}\sim
t^{\threehalves}$, and $x_{\rm typ}\sim\,t^{5/2}$, respectively.
We proceed as we did for the $n=2$ case.
Once again, we divide the time axis by the zero-crossings
of the velocity. The time intervals $\Delta t_i$ are now distributed with
the Levy exponent $\beta_2=\fourth$.
The sum of $N$ independent $\Delta t$'s is approximately zero for $t<N^4$, and 
for larger $t$, is approximated by
\begin{equation}
L_N(t)dt\sim \frac{N}{t^{\fourth}}\frac{dt}{t}.
\end{equation}

Proceeding casually, we may think that, as before,
the distribution $f(N)$ has Levy exponent of 
$\half$ since that result is universal for all symmetric
walks.  We perform an integral analogous to Eq. (14)
and obtain a first passage time Levy exponent $\beta_3=\eighth$.
However, our numerical simulations yield a Levy exponent of $0.220\pm0.001$.
The exponent is closer to $\fourth$ than $\eighth$.  Clearly,
there is something wrong.

The method fails because there are now \emph{correlations} 
between the $\Delta x_i$ steps, where $\Delta x_i$ is the change 
in $x$ during a $\Delta t_i$ step.  
In the previous $n=2$ case, the velocity undergoes a simple 
random walk and so its sign is random after a zero-crossing.  
In this $n=3$ case, the velocity is the integral of a
random walk.  It is a differentiable function and so almost
certainly changes sign when it hits zero.  Consequently,
the $\Delta x_i$ steps alternate in sign.
In addition, as the acceleration drifts about, the $\Delta x_i$ steps
increase in size. The acceleration at the beginning of each next time
interval is larger on average than the previous one, making the next
$\Delta t_i$ and, therefore, the next $\Delta x_i$ larger
than the last.  

A large number of alternating, increasing steps will
reach zero quickly. From the numerical simulation, 
with each oscillation there is
a definite probability of reaching zero which is almost constant.  In other
words, the
distribution $f(N)$ is not a power law as in the $n=2$ case, but an
exponential,
\begin{equation}
f(N)\sim  e^{-\lambda_{3} (2N+1)}\ \ \ \ {\rm with}\ \lambda_{3}=
0.79\pm\,0.02.
\end{equation} 
See Table II and the corresponding Fig. 2.  The number of velocity
zero-crossings must be odd because the position can become negative 
only after the velocity has become negative. 
With this exponential distribution, 
the number of time intervals one must add up is actually quite
small.  However, adding a finite number of 
independently drawn Levy variables distributed
with Levy exponent $\fourth$ produces a flight with Levy exponent 
$\fourth$, so it is quite surprising that adding $N$ of them together,
where $N$ has a finite mean, leads to anything other than an 
exponent of $\fourth$.

And yet, adding a small number of \emph{correlated} Levy variables
does shift the exponent.  To quantitatively describe 
the correlation in magnitude, we need to determine
how the initial acceleration sets the time scale for each 
$\Delta t_i$ interval.  
We saw from Eq. (3) for the $n=1$ random walk that  
the time scale
is the initial value squared.  Therefore, the time scale 
between two acceleration zero-crossings
is the initial acceleration squared.
Because the time between velocity zero-crossings
is comprised of many acceleration zero-crossings, the time scale
between velocity zero-crossings 
is fixed in the same way.  

This argument is somewhat general, so it is nice to verify that it
is correct. The coefficient in the formula for the first passage time
distribution was computed in Ref. [2] for the case of the $n=2$ random walk
by solving the Fokker-Planck equation with the appropriate boundary 
conditions.
\begin{equation}
\rho_{2}(t;v_{o},x_{o}\rightarrow\,0)dt \sim \, \frac{\sqrt{v_o}}{t^{1/4}}
\frac{dt}{t} = (\frac{v_o}{\sqrt{t}})^\half \frac{dt}{t}
\end{equation}
for large t. Note that $t_{\rm typ}\sim v_o^2$. Since the acceleration
in the $n=3$ case is exactly the same as the velocity in the $n=2$ case,
we can translate this result into $n=3$ language by replacing $v$ with $a$. 
Therefore
in the $n=3$ case, the square of the initial acceleration determines the
time scale until the next zero-crossing. 

So the square of the initial
acceleration at each time step determines the scale of the next time step.
Since the acceleration is an ordinary random walk, its square has a size
proportional to the total elapsed time. Therefore, $\Delta t =
T\times q $ where $q$ is a 
unit Levy variable and $T$ is the total elapsed time.
To describe this 
correlated process, we set
the units of time so that  $\Delta t_1$
is a unit size Levy variable $q_1$. The next time step, $\Delta t_2$, is no
longer unit sized, but has a magnitude determined by the square of the
acceleration, or, equivalently, the total elapsed time. So to find
$\Delta t_2$ we
take $\Delta t_1$ and multiply it by separate, independent, unit Levy variable
$q_2$.
To find $\Delta t_3$, we take the total elapsed time, $\Delta t_1 + 
\Delta t_2$, which is the expected square acceleration,
and multiply by $q_3$. In equations, 
\begin{equation}
\Delta t_1 = q_1 
\end{equation}
\[
\Delta t_2 = (\Delta t_1) q_2 \]\[
\Delta t_3 = (\Delta t_1 + \Delta t_2) q_3 \]\[
\Delta t_4 = (\Delta t_1 + \Delta t_2 + \Delta t_3) q_4 \ .\]
The total time $t$ that has elapsed after $N$ steps is then
\begin{equation}
t = q_1 \prod_{i=2}^N (1+q_i) \ .
\end{equation}
 
The asymptotic distribution is extracted when $q_i>>1$. This means we
need to determine the distribution of products of Levy variables.
Taking the natural logarithm, the product becomes a sum. Define
$z_2=ln(\Delta t_2)$, $f_1=ln(q_1)$, and $f_2=ln(q_2)$.
We use the following approximate distribution for both Levy
variables,
\begin{equation}
P(q)dq = {\alpha c_0^{\alpha} \over q^{\alpha}}\frac{dq}{q}
\end{equation}\[
P(q)dq = 0\ \ \ q<c_0 .
\]
with $c_0=1$. Keep in mind, in order to 
compute $\beta_3$, the Levy exponent $\alpha$
of the $q_i$ variables equals $\fourth$.  
In terms of the transformed variables, the distribution 
of $f_1$ and $f_2$ is given by
\begin{equation}
P(f)=\alpha\,e^{-\alpha\,f} \ .
\end{equation}
If $q_1$ and $q_2$ had {\em different} Levy exponents, $\alpha_1>\alpha_2$,
the distribution of the sum would be 
\begin{equation}
\int_0^{z_2}df_1 P(f_1)P(z_2-f_1)=
\frac{\alpha_1\alpha_2}{(\alpha_2-\alpha_1)}(e^{-\alpha_1\,z_2}-
e^{-\alpha_2\,z_2})\ .
\end{equation}
For large $z_2$, $P(z_2)\approx \alpha_2\,e^{-\alpha_2 z_2}$, i.e.
the smaller exponent dominates the tail.
However, when $\alpha_1=\alpha_2=\alpha$,
we have
\begin{equation}
P(z_2)=\alpha^2 z_2 e^{-\alpha z_2} \ .
\end{equation}
In terms of $q_1$ and $q_2$, the probability distribution of $q_1 q_2$
is [8]
\begin{equation}
P(\Delta t_2)d(\Delta t_2)=
\alpha^{2}\frac{ln(\Delta t_{2})}{\Delta t_{2}^{\alpha}}\frac{d(
\Delta t_2)}
{\Delta t_2} \ .
\end{equation}
In general, the probability distribution for $i$ independent Levy 
variables multiplied together is
\begin{equation}
P(z_i)=\frac{\alpha^{i}}{(i-1)!}z_i^{i-1}e^{-\alpha z_i} \ .
\end{equation}
The $(i-1)!$ is required for normalization. Transforming back to the
original $\Delta t_i$ variables,
\begin{equation}
P(\Delta t_i)d(\Delta t_i)=
{\alpha^i \over (i-1)!}
{(ln \Delta t_i )^{i-1} \over \Delta t_i^{\alpha}}\frac{d(\Delta t_i)}
{\Delta t_i}.
\end{equation}  
We substitute $i=N$ and $\alpha=\fourth$ and this distribution becomes
$L_N(t)$--- the probability distribution of the
$N$th passage time of the velocity.
Unlike the $N$th passage time of the
acceleration, $L_N(t)$ is not a Levy stable distribution because of the 
correlations.  More specifically, $L_N(t)$ is the distribution of a 
product of $N$ Levy variables $\prod q_i$.
The expression for the $N$th passage time we derived earlier
is $\prod (q_i + 1)$, which differs in subleading terms from the 
previous expression.

Note that $L_N(t)$ coincides with the $N$th passage time
distribution for the position in the $n=2$ case. Both the velocity for
$n=3$ and the position for $n=2$ are integrals of an $n=1$ random walk. 
Observe that the distribution $L_2(t)$ is {\em not} 
the same as the distribution of the sum
of two variables distributed as $L_1(t)$--- there is an additional logarithm.
We determined $L_2(t)$ numerically and it agrees with this form (Fig. 3).
For higher $N$, there are $N-1$ factors of $ln(t)$.  

Before estimating $\beta_3$, we must make one last argument in the
form of an approximation.  Up until this
point, $L_N(t)dt$ assumes that each step
is free to be larger than the previous one.  This is 
because we have not yet taken
into account the conditional dependence of the correlation on the position
being positive for the walk to persist.  In 
fact, the generic situation 
when the steps are ever-increasing and alternating in
sign is for there to be only one step.  This means that walks 
that survive more
than one step are special.  For these walks, the 
second step is an exceptional pick from the
distribution.  The \emph{easiest} way for the walk to survive is
for the second step to be approximately the same size as
the first step.  In other words, the size of the 
first step cuts off the distribution
of the second step.  The result is that the second step shares the
distribution of the first.  For the third step, the position is positive,
and now its time interval is
distributed the same as the second step would have been.  
Generalizing our argument,
any two \emph{complete} time intervals can simply be lumped 
together as one and
we conclude that the distribution
of the total after $N$ steps is the same as the distribution of the
step sizes after $N \over 2$ steps.  In other words, $L_N(t)dt$ is not 
the correct distribution for $N$ steps, but $L_{N/2}(t)dt$ is.
We present evidence for the even-numbered complete steps being typically small 
in Fig. 4.  For $n=3$, the second
step is, on average, about five times smaller than
the first time interval.  For $n=4$, the second time interval
is typically twice as small as the first.  

So we now have all the ingredients to estimate the Levy
exponent $\beta_3$.  We arrive at
\begin{equation}
\rho_{3}(t)dt\sim e^{-\lambda_3}\frac{1}{t^{5/4}}\sum_{N=0}^
{\infty}\,\frac{1}{4^{N}}\frac{1}{N!}(ln(t))^{N}e^{
-2\lambda_{3}\,N}dt
\end{equation}
\begin{equation}
=e^{-\lambda_3}\frac{1}{t^{5/4}}e^{\frac{e^{-2\lambda_{3}}}{4}ln(t)}dt,
\end{equation}
which gives $\beta_3^{{\rm est}}=\fourth - {e^{-2\lambda_3} \over 4}$.
Substituting the numerical value $\lambda_3=0.79$,
we find the Levy index $0.199$.  This result 
is a shift in the right direction from our original estimate, 
but it is still too large.  

Figure 4 suggests that our estimation scheme may work better
for $n>3$ and so we extend the method to $n=4,5,...$.  
As before, we define the two Levy flights of the time and position intervals
between velocity zero-crossings. 
For $n>3$, the size correlation factor for the time interval Levy flight 
turns out to be the same factor of the elapsed time as in the $n=3$ case 
with each step 
retaining only the memory of the highest-order derivative.  To be
more precise, the initial
acceleration at the start of each time interval for the $n=3$ case 
is replaced with the 
highest-order
random variable.  We ignore any other memory effects.  Then, for the $n$th-order process, the step-size distribution for the
$\Delta t_i$ intervals are governed by the passage time exponent for the
$(n-1)$th-order process.  Referring back to Eq. (28), $i=N$ and 
$\alpha=\beta_{n-1}$.  In addition, 
the correlation 
in the sign of the $\Delta x_i$ 
still persists for the $n$th-order random walk so the
distribution $f(N)$ remains exponential with a decay constant that
we determine numerically.  Therefore, the $n{\rm th}$ Levy exponent will
be a small perturbation about the $(n-1)$th first passage time exponent with
a shift of $-\beta_{n-1}e^{-2\lambda_{n}}$.   
For $n=4$ and $n=5$, we observe that our estimation scheme yields more 
accurate results.  See Table II.  

As in the $n=2$ case, for $n\geq 3$ we consider the
last partial time step to be a full one.  Given that $L_{N/2}(t)dt$ 
is the more appropriate distribution, the bulk of the first passage
time for a given $N$ is taken up during the last two steps (up to 
subleading corrections).   So the these last partial time steps 
should be distributed just as the first passage time.
We have numerically verified this to be the case.

Majumdar {\it et. al}. [5] have constructed a
completely different approximation scheme to compute the
first passage time exponent for these higher-order random
walks.  While we refer the reader to Ref. [5] for the details, their scheme
is rather accurate for $\beta_2$ and $\beta_3$.
 
\section{The $\lowercase{n}=\infty$ Limit}

As $n$ increases, we also expect the 
decay constant $\lambda_n$ to grow.  
In fact, in the limit that $n$ becomes large,
the first passage time exponent reaches a limiting value.  
We consider the solution to Eq. (1) with initial conditions
$x(0)=x'(0)=...=x^{n}(0)=0$.  It is
\begin{equation}
x(t)=\int_0^t dt_n \int_{0}^{t_n} dt_{n-1} ... \int_0^{t_2} dt_1 
\eta(t_1),
\end{equation}
which is a linear function of $\eta$ and therefore is 
the inner product of $\eta$ with a kernel.  In this case,
\begin{equation}
x(t)=\int_0^t dt' {(t-t')^{(n-1)} \over (n-1)!}  \eta(t') . 
\end{equation}
It is easy to see that this expression satisfies Eq. (1).

When solving for the time when $x(t)$ first 
becomes zero, it is permitted to rescale
$x(t)$ by any finite number, even if that number is a function of time. This
does not change the set of points where $x(t)$ is zero. Rescaling by 
$t^n$ so that $y(t)=\frac{x(t)}{t^n}$ and dropping constant factors,
\begin{equation}
y(t)=\frac{1}{t}\int_0^t dt' (1-{t'\over t})^{n-1} \eta(t') \ .
\end{equation}

In the large $n$ limit, $(1-\frac{t'}{t})^{n}$ is 
indistinguishable from an exponential in
the region where it has the most weight ($t' < nt$). The result is that
\begin{equation}
y(t)=\frac{1}{t}\int_0^t dt' e^{-nt'\over t} \eta(t') \ .
\end{equation}

Majumdar {it et al}. [5] have noted that the $n=\infty$ random walk has the
same first passage time exponent as a very different system --- the
solution of the two-dimensional diffusion equation with
a random initial condition,
\begin{equation}
{d\rho\over dt}=\nabla^2 \rho 
\end{equation}
\[
\rho(x,y,0)=\eta(x,y) \ .
\]
In polar coordinates, the value of $\rho(0,t)$ is given by
\begin{equation}
\rho(0,t)=\frac{1}{2\pi t}\int d^2\vec{r} \gaussian{r}{2t} \frac{1}{\sqrt{r}}
\eta(r,\theta) \ ,
\end{equation}
where 
we have used $\delta(x)\delta(y)=
r^{-1}\delta(r)\delta(\theta)$.
The angular integration leaves the integral in terms of a new random variable,
$\eta'(r)= {1\over \sqrt{2\pi}}\int d\theta \eta(r,\theta)$.  We now have
\begin{equation}
\rho(0,t) = \frac{1}{\sqrt{2\pi}t}\int dr \sqrt{r} \gaussian{r}{2t} \eta'(r)
\end{equation}
Transforming variables to $u=r^2$, we arrive at
\begin{equation}
\frac{1}{2}\frac{1}{t\sqrt{\pi}}\int du e^{-\frac{u}{2t}}\eta(u),
\end{equation}
where we used the fact that $\delta(r)=\delta(u)2\sqrt{u}$ so that
$\eta'(r)=\eta(u)\sqrt{2}u^{\frac{1}{4}}$.

Comparing this kernel with the kernel for
the $n=\infty$ random walk, we see that they are indeed the same.
As a result, the distribution of times for the diffusive field to
change sign at one point is the same as the distribution of the
first passage time for the $n=\infty$ random walk for long times,
where the initial conditions are irrelevant.

\section{Physical Systems}

Consider a planar crack front moving
through rough, brittle material.  In the quasistatic 
regime, the planar crack
front is a line with long-range static 
elasticity.  There is a uniform stress applied
to the material, which drives the crack front forward.  But there are
also local, random variations in the fracture toughness which the crack
front needs to overcome to move forward.   These randomly varying forces are 
pinning forces.  The competition 
between the global pulling and local pinning forces as mediated by the
elasticity of the crack front determines its dynamics.  For small values
of the applied force, the pinning forces eventually 
dominate and the crack front
remains stuck.   For some 
finite value of the applied force, the pinning forces are not strong enough
to keep the crack front in place and it moves forward.  The
boundary between these two phases is
the depinning transition.

Approaching the depinning transition from the static side, a small increase
in the applied force causes a tiny portion
of the crack front to move forward.
This portion causes a few
more regions to move forward as well, only to be eventually 
stopped by more strongly
pinned regions. The only role of the applied force here is to induce the
initial motion of the tiny portion.  It remains fixed until the local
motion of the crack front stops.   Near the 
depinning transition, there exists a sequence of these discrete, localized
\emph{avalanche events} with distribution of sizes $\tau$.  The  
avalanche
size is defined as the total amount of the crack front that has
moved forward during these events.  A power law distribution of avalanche
sizes as the force approaches the critical force 
indicates a continuous, second-order-like depinning transition.

Near the depinning transition, the motion of the crack front is
jerky.  We therefore model its dynamics in terms of 
discrete space and time.  Within an infinite-range model, where 
each crack front segment is coupled
equally to every other, the spatial degrees of freedom along the crack front
are averaged out.  We define $y_t$ 
to be the total number of segments of the crack front that hop
forward at discrete time $t$.  Given that we are only concerned with the few
segments that are on the verge of hopping forward, $y_t$ is Poissonian
distributed.  In this infinite-range 
limit, and in the dissipative regime,  
the mean of $y_t$ is determined by the 
number of segments that have hopped at the 
previous time step only.  In addition, the mean of $y_t$ 
fluctuates statistically until it reaches zero, whereupon the motion
stops.  In the limit of large avalanches where things are changing slowly with
time, the equation of motion 
for the avalanches at the depinning transition is
\begin{equation}
{dy \over dt}=y^{\half} \eta(t) \ 
\end{equation}
in the Ito representation (the time derivatives are forward differences).
When $y(t)$ reaches zero, the avalanche is over.   We refer the reader
to Ref. [6] for a more detailed derivation.
  
So far, we have neglected any effects like elastic waves that are 
indeed present along the crack front.  If a piece of the crack front moves
forward, that motion creates an extra transient force on its neighbors as
the elastic wave propagates along the crack front.  The extra transient
stress is called a stress 
overshoot.  This effect
is different than pure inertia, which would be an extra transient
stress on the segment itself after it jumps.  However, we demonstrate that
the two effects are similar in this infinite-range model for a particular 
value of the stress overshoot [6].     
If we were to take into account elastic waves along the 
crack front, the mean of $y_t$ depends on several previous time steps.
In other words, there 
are higher-order corrections to 
Eq. (39) that are usually irrelevant in the long time limit.  The most
significant terms are the first and second time derivatives, and
the first time derivative is the only relevant term.
 If we fine-tune the coefficient of
${dy \over dt}$ to zero, we arrive at a tricritical point.  Here, the
second-order derivative is the most relevant term and we obtain the
equation of motion
\begin{equation}
{d^2y \over dt^2}=y^\half \eta(t) .
\end{equation}
The variable we are interested in is
\begin{equation}
\tau=\int_{0}^{t_f}dt\,y(t)\, , 
\end{equation}
the avalanche size.  We emphasize that $t_f$ is determined by the boundary 
condition $y(t_f)=0$.

In order to find the distribution of $\tau$, we use the size of the
avalanche as a new path-dependent time coordinate.
We transform time on each particular path in the path-integral differently,
in a way that depends on the history of $y(t)$. This
transformation alters the density of $y(t)$'s at any given time,
mixing together $y(t)$'s from different times so that they appear simultaneous.
Because of this property, it is difficult to perform the transformation in the
Fokker-Planck equation; but the transformation is easy
in the path integral.

A reparameterization of time, even a path
dependent one, does not affect the answers to questions that do
not involve the time explicitly. 
The probability for a stochastic walker to arrive at point B
from point A is independent of the global time, but the probability
for a stochastic walker to be at point B at time $t$ is not.
Similarly, the avalanche size of a given walk does not involve the
time lapse of the walk as the upper cutoff of the integral is
determined by the boundary condition on $y(t)$ and not the time
explicitly.

Choosing the new time coordinate $\tau$ to tick at the new rate,
\begin{equation}
{d\tau \over dt}=y,
\end{equation}
and using the property of Gaussian noise,
$\eta(\tau(t))=\frac{\eta(t)}{\sqrt{\frac{d\tau}{dt}}}$, 
the (Ito) equation pair 
\begin{equation}
{dv\over dt}=y^{\half}\eta(t)
\end{equation}
\[
{dy\over dt}=v
\]
become
\begin{equation}
{dv\over d\tau}=\eta(t)
\end{equation}
\[
y{dy\over d\tau}= v.
\]
Replacing $y$ with $x=\frac{y^2}{2}$ completes the transformation.  
The zeros of $y$ are also the zeros of $x$ so the value 
of $\tau$ when $x$ is
zero is then the avalanche size.  
We therefore find the
distribution of avalanche sizes to be the distribution of first
passage times $\tau$ for the $n=2$ random walk. It has a Levy exponent
$\fourth$.

We may perform the same transformation on the Fokker-Planck equation directly,
although the motivation then becomes obscure. Consider the 
following probability-conserving equation,
\begin{equation}
\frac{\partial P}{\partial t} =  -y
\frac{\partial P}{\partial \tau}
+\half y\frac{\partial^2 P}{\partial v^2} - 
v\frac{\partial P}{\partial y},
\end{equation}
for a new quantity $P_t(x,v,\tau)$. If we integrate $P_t(x,v,\tau)$ 
over all $\tau$,
we recover the original Fokker-Planck equation. On the other hand, if
integrate $P_t(x,v,\tau)$ 
with respect to $t$ from zero to infinity, we obtain
\begin{equation}
 \frac{\partial K}{\partial \tau}
=\half \frac{\partial^2 K}{\partial v^{2}}- 
\frac{v}{y}\frac{\partial K}
{\partial y},
\end{equation}
where $K(v,y,\tau)=\int_{0}^{\infty}dt\,P_t(v,y,\tau)$. 
Defining $x=\frac{y^2}{2}$, as before,
gives
\begin{equation}
 \frac{\partial K}{\partial \tau}
=\half \frac{\partial^2 K}{\partial v^{2}}-v\frac{\partial K}{\partial x},
\end{equation} 
which is the Fokker-Planck equation for the $n=2$ random walk. This demonstrates
the equivalence of the two problems.

It should now be clear that the avalanche size exponent
for Eq. (39) is $\frac{1}{2}$ since the same time reparameterization 
may be performed in that case also.  This relates the avalanche size 
exponent to 
the first passage
time of the $n=1$ random walk.

\section{Conclusions}

We have presented a relation between the shift in the first passage time
exponent and the decay rate of the probability of $N$ velocity zero-crossings
for the $n$th random walk.  The method
can be exact for $n=2$, 
but not for $n \geq 3$ because of correlations.  More work is needed to 
estimate the decay constant of
$f(N)$ since this would determine the
convergence rate of the exponents to the $n=\infty$ value. 
In addition, we do not have any bounds on the error of the first passage time
exponent as the approximation is uncontrolled.  
We also do not know which types of correlations for Levy flights 
lead to shifts in exponents and which do not.  A classification of 
correlations in terms of these two categories might be useful.

We have opted to slice up time in 
terms of velocity zero-crossings, instead 
of using the global time in the Fokker-Planck equation.  This approach
allows us to analyze the higher-order random walks in terms of a
one-dimensional Levy process since the phase variables are
the position and the time only. The remaining variables are subsumed in
the $(n-1)$th-order random walk that the velocity undergoes between
the zero-crossings.
 
With Eq. (42) we have also reparameterized the
global time, but in a different way. We use a path-dependent time
transformation to analyze the avalanche statistics of a nonlinear
second-order random walk in terms of the first passage time exponent
of the linear second-order random walk. This allows us to give a novel physical
interpretation of the first passage time exponent for the $n=2$ case.

Since the avalanche size does not depend on the global time, it has been
known to us that one 
can find the stationary solution of the Fokker-Planck equation corresponding
to the nonlinear random walk in Eq. (40) and then arrive at the
avalanche size exponent after invoking a simple scaling argument [9].  
Since this method gives a
new derivation of the avalanche size exponent, it indirectly gives a
new derivation of the
first passage time exponent for the $n=2$ case.  This insight may be
useful to construct a derivation of $\beta_n$ for $n\geq 3$.

Finally, these higher-order random walks could be relevant for more
general forms of dynamic stress transfer along a crack front, 
or other extended elastic objects. For instance, one can approach a fine-tuned
critical point where the coefficients of the first and second time derivative
terms are zero. Then the third-order time derivative is the
most relevant term, and we arrive at a higher-order critical point.  
While the time reparameterization given by Eq. (42) 
does not allow us to equate the avalanche
size exponent with the first passage time exponent in this case, we remain optimistic that either exponent may be relevant for various physical systems.

\section*{Acknowledgements}
The authors would
like to thank P.A. Argyres and Lev Kaplan for useful 
discussion.  J.M.S. would also especially like to thank Daniel S. Fisher for 
suggesting the problem, for many helpful discussions and for useful comments
on the manuscript.  
J.M.S. has been supported by the National Science Foundation
via DMR-9630064, DMR-9976621, and DMR-9809363.  R.M. has been
supported by PHY95-13717.  

\appendix
\section*{The Kernel Of Higher-Order Random Walks}

We begin with the partition function for all higher-order random walks,
\begin{equation}
Z=\int\,D[\eta(t)]\,e^{-\frac{1}{2}\int_{0}^{T}(\eta(t))^{2}}\,.
\end{equation}
We change variables from $\eta(t)$ to $x(t)$, 
where ${d^n x \over dt^n}= \eta(t)$, and
use forward difference time derivatives. We change variables with the insertion 
$\int_{x(0)=x_o}^{x(T)=x_f}\,D[x(t)]\delta(\frac{d^{n}x(t)}{dt^{n}}-\eta
(t))$ to arrive at
\begin{equation}
\int_{x(0)=x_o}
^{x(T)=x_f}D[x(t)]e^{-\half \int_{0}^{T} (\frac{d^{n}x(t)}{dt^{n}})^2}.
\end{equation}
We then minimize the action to find the classical equation of motion
with vanishing variations on the boundary for all derivatives up to the
$(n-1)$th-derivative.  In the $n=2$ case we find the equation of motion,
\begin{equation}
\frac{d^4 x(t)}{dt^4}=0.
\end{equation}
We 
then impose the following constraints on the $x(t)$ trajectory,
$x(0)=x_o$, $x(T)=x_f$, $x'(0)=v_o$, and $x'(T)=v_f$, yielding 
\begin{eqnarray}
x(t)=x_{o}+v_{o}t+\frac{(3x_{f}-3x_{o}-v_{f}T-2v_{o}T)}{T^{2}}t^{2} \nonumber \\
+\frac{(v_{f}+v_{o}-\frac{2}{T}(x_{f}-x_{o}))}{T^{2}}t^{3}\,.
\end{eqnarray}

The path integral is quadratic, so the partition function is proportional
to the exponential of the classical action.  Substituting the classical
solution into the action and integrating gives the propagation
kernel,
\begin{eqnarray}
P_{t}(v_{f},x_{f};v_{o},x_{o})= 
\frac{\sqrt{3}}{\pi\,t^{2}}\,e^{-\frac{(v_{f}-v_{o})^{2}}{2 t}} \nonumber \\
e^{-\frac{6((x_{f}-x_{o})-\frac{1}{2}(v_{f}+v_{o})t)^{2}}{t^{3}}} \,.
\end{eqnarray}
We normalize the propagation kernel 
to have integral unity.  Notice that the kernel is the product of
a Gaussian in $v$ and a gaussian in $x$. Given that $x(t)$ is the time
integral of $v(t)$ and that the sum of Gaussian variables is
Gaussian, if the velocity is Gaussian, so is the position.
Not only does the position spread faster than the velocity,
but it also drifts with time with a coefficient that depends on the 
velocity.
This method works for higher $n$ giving us the following 
propagation kernels for $n=3,4,5$:

\begin{eqnarray}
P_{t}(a_f,v_f,x_f;a_o,v_o,x_o)\!\sim\!\frac{1}{t^{\frac{9}{2}}}
e^{-\frac{1}{2 t}(a_f-a_o)^2} 
e^{-\frac{6}{t^{3}}((v_f-v_o)-\frac{1}{2}
(a_f+a_o)t)^{2}}                        \nonumber \\
e^{-\frac{360}{t^{5}}((x_f-x_o)-\frac{1}{2}(v_o+v_f)t
+\frac{1}{12}(a_f-a_o)t^{2})^2}
\end{eqnarray}
for $n=3$,
\begin{eqnarray}
P_t(s_f,a_f,v_f,x_f;s_o,a_o,v_o,x_o)\sim\frac{1}{t^{8}}
e^{-\frac{(s_f-s_o)^{2}}{2t}} \nonumber \\
e^{-\frac{6}{t^{3}}((a_f-a_o)-\frac{1}{2}(s_f+s_o)t)^{2}}   \nonumber \\
e^{-\frac{360}{t^{5}}((v_f-v_o)-\frac{1}{2}
(a_f+a_o)t+\frac{1}{12}(s_f-s_o)t^{2})^{2}} \nonumber \\
e^{-\frac{50400}{t^{7}}((x_f-x_o)-\frac{1}{2}(v_f+v_o)t+\frac{1}
{10}(a_f-a_o)t^{2}-\frac{1}{120}(s_f+s_o)t^{3})^{2}}
\end{eqnarray}

for $n=4$ with $\frac{da(t)}{dt}=s(t)$, and
\begin{eqnarray}
P_t(u_f,s_f,a_f,v_f,x_f;u_o,s_o,a_o,v_o,x_o)\sim\frac{1}{
t^{\frac{25}{2}}}e^{-\frac{(u_f-u_o)^{2}}{2 t}}
e^{-\frac{6}{t^{3}}((s_f-s_o)-\frac{1}{2}(u_f+u_o)t)^{2}}  \nonumber \\
e^{-\frac{360}{t^{5}}((a_f-a_o)-\frac{1}{2}
(s_f+s_o)t+\frac{1}{12}(u_f-u_o)t^{2})^{2}}
e^{-\frac{50400}{t^{7}}((v_f-v_o)-\frac{1}{2}(a_f+a_o)t+\frac{1}
{10}(s_f-s_o)t^{2}-\frac{1}{120}(u_f+u_o)t^{3})^{2}} \nonumber \\
e^{-\frac{12700800}{t^{9}}((x_f-x_o)
-\frac{1}{2}(v_f+v_o)t+\frac{3}{28}(a_f-a_o)t^{2}-\frac{1}{84}
(s_f+s_o)t^{3}+\frac{1}{1680}(u_f-u_o)t^{4})^
{2}}
\end{eqnarray} 
for $n=5$ with $\frac{ds(t)}{dt}=u(t)$.


\begin{table}
\caption{The first passage time exponents obtained from fits to
the data shown in Fig. 1.}
\begin{tabular}{ccc}   
$n$ & $\chi$  & $\beta_{n}$  \\ 
\hline
2  &0.005    &0.250$\pm$0.001    \\
3  &0.005    &0.220$\pm$0.001    \\
4  &0.005    &0.210$\pm$0.001    \\
5  &0.005    &0.204$\pm$0.001    \\
\end{tabular}
\end{table}

\begin{table}
\caption{The first passage time exponent estimates for 
the $n\ge\,3$, which uses the numerical
exponential decay constant $\lambda_{n}$ and the 
numerical results for $\beta_{n-1}$.  This estimate is to be compared with
the last column in Table I.}
\begin{tabular}{cccc}   
$n$ & $\chi$  & $\lambda_{n}$ & $\beta_{n}^{{\rm est}}$ \\ 
\hline
3  &0.005   &0.79$\pm$0.02     &0.199         \\
4  &0.005   &1.55$\pm$0.02   &0.210         \\
5  &0.005   &1.92$\pm$0.02   &0.205         \\ 
\end{tabular}
\end{table}

\begin{figure}[h]
\begin{center}
\epsfxsize=7.5cm
\epsfysize=7.5cm
\epsfbox{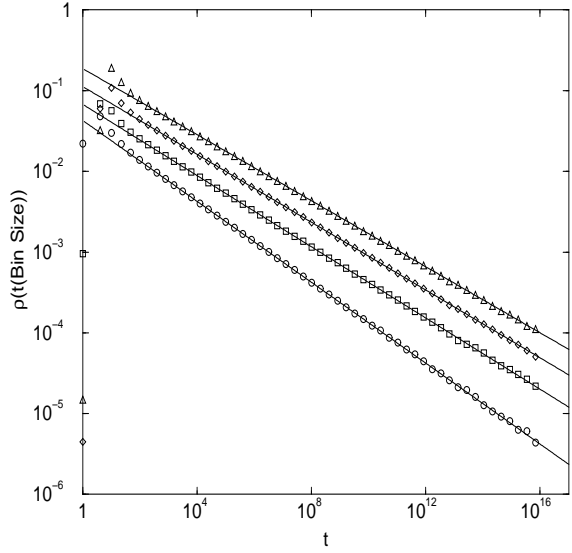}
\caption{Log-log plot of the probability of a first passage time $t$ 
(in arbitrary units) occurring within 
the interval [$t/\sqrt{2},\sqrt{2}t$), where the probability has been
multiplied by a different constant for each curve so that they do 
not overlap.  
The open circles
denote numerical data for $n=2$; the open squares for $n=3$; the open diamonds for 
$n=4$;
the open triangles for $n=5$.  The solid lines represent the results of 
the linear regression.  The size of the symbols
is larger than the error bars.  As we only used
double precision in our simulations, there is an upper cutoff in the
first passage time of approximately $10^{16}$.}
\end{center}
\end{figure}

\vspace*{1.0cm}
\begin{figure}[h]
\begin{center}
\epsfxsize=7.5cm
\epsfysize=7.5cm
\epsfbox{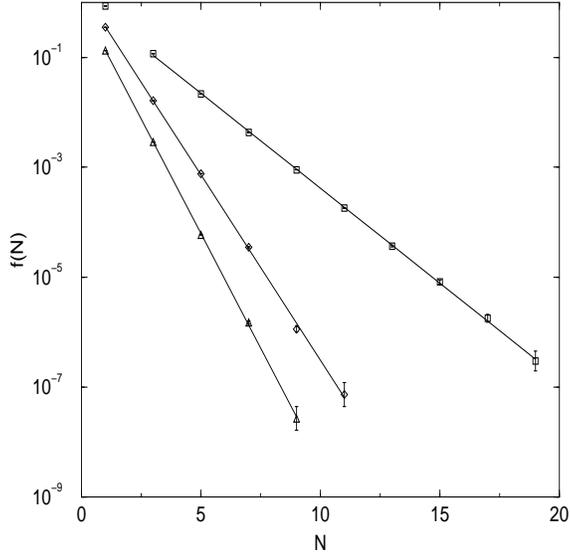}
\caption{Log-linear plot of the distribution of 
number of velocity zero-crossings $f(N)$, where $f(N)$ has been multiplied by a 
different constant for each curve so that they shift and do not overlap.  
The symbols here
are the same as in Fig. 1.  For the 
$n=3$ curve, we ignore the first data point
in the linear regression as there is some memory of the initial conditions
for this point.}
\end{center}
\end{figure}

\begin{figure}[h]
\begin{center}
\epsfxsize=7.5cm
\epsfysize=7.5cm
\epsfbox{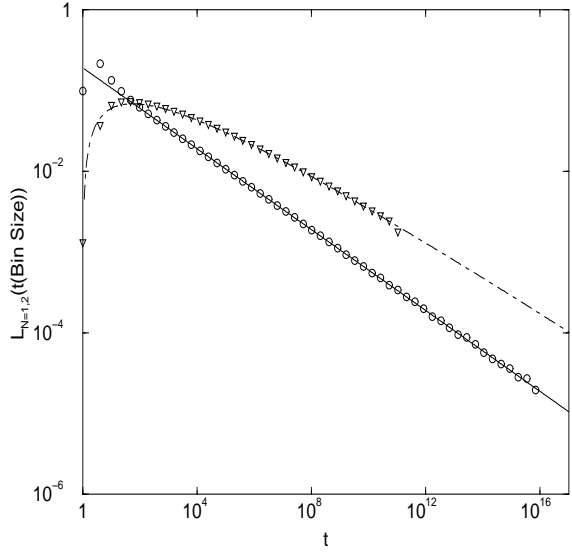}
\caption{Log-log plot of $L_1(t)$ and $L_2(t)$ 
for the $n=2$ case, i.e. the probability of a first 
(or second) position passage
time $t$ (in arbitrary units) occurring 
within the interval $[t/\sqrt{2},\sqrt{2}t)$.    The open circles denote numerical data for $L_1(t)$; the inverted triangles for 
$L_2(t)$.  While
the solid line represents the results of the linear regression; 
the dot-dashed line represents Eq. (28) up to a 
proportionality constant, with $i=2$ and $\Delta t_i=t$.  
As we must keep track of even larger absolute 
values of the position for $N=2$, the double precision constraint
cuts off the tail of the distribution even more so for $N=2$ than for $N=1$.
Once again, the symbols are larger than the error bars and there are 
approximately $10^7$ samples.
}
\end{center}
\end{figure}

\begin{figure}[h]
\begin{center}
\epsfxsize=7.5cm
\epsfysize=7.5cm
\epsfbox{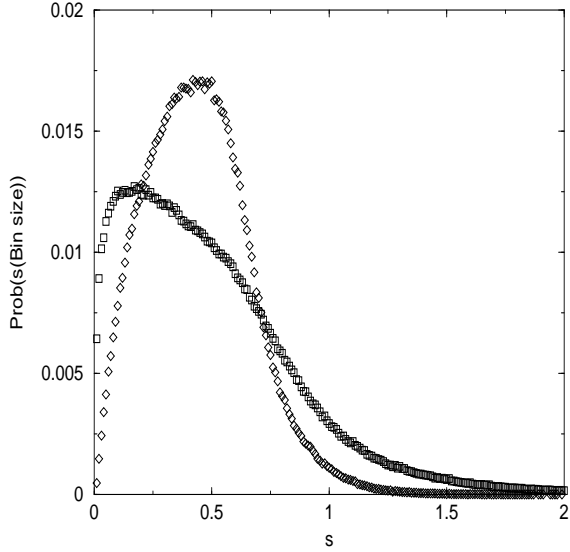}
\caption{Plot of the probability of the ratio of the 
second complete time interval
to the first complete time interval, denoted as $s$, 
occurring within the interval [$s,s+0.01$) for the $n=3$ (the open squares) 
and $n=4$ (the open diamonds) 
random walk.  
Both probabilities  
eventually fall off exponentially
as opposed to being distributed as $L_2(\Delta t_2)d(\Delta t_2)$,
indicating that the second step is indeed an exceptional pick from 
the distribution for the first time interval.  Each curve contains
approximately $10^6$ samples. }
\end{center}
\end{figure}

\end{document}